\documentclass[twocolumn, tighten]{aastex63}

\usepackage{nicefrac}
\usepackage{microtype}
\usepackage{booktabs}
\usepackage{rotating}
\usepackage{multirow}

\newcommand{\lya}{\ensuremath{\rm{Ly}\alpha}}
\newcommand{\lyb}{\ensuremath{\rm{Ly}\beta}}

\newcommand{\kms}{\ensuremath{{\,\,\rm km\,s}^{-1}}}

\newcommand{\flux}{\ensuremath{\,\,{\rm erg}\,\,{\rm cm}^{-2}\,\,{\rm s}^{-1}}}
\newcommand{\lum}{\ensuremath{\,\,{\rm erg}\,\,{\rm s}^{-1}}}

\newcommand{\cloudy}{\textsc{Cloudy}}
\newcommand{\calcos}{\textsc{calcos}}

\newcommand{\JS}[1]{{\color{black}#1}}
\newcommand{\SDJ}[1]{{\color{black}#1}}
\newcommand{\AG}[1]{{\color{black}#1}}
\received{}
\revised{}
\accepted{}

\submitjournal{ApJL}
\shorttitle{}

\begin{document}

\title{Spatially-resolved UV diagnostics of AGN feedback: radiation pressure dominates in a prototypical quasar-driven superwind}
\author[0000-0001-8426-5732]{Jean Somalwar}
\affil{Department of Astrophysical Sciences, 4 Ivy Lane, Princeton University, Princeton, NJ 08544, USA}

\author[0000-0001-9487-8583]{Sean D. Johnson}
\altaffiliation{Carnegie-Princeton fellow}
\affil{Department of Astrophysical Sciences, 4 Ivy Lane, Princeton University, Princeton, NJ 08544, USA}
\affil{The Observatories of the Carnegie Institution for Science, 813 Santa Barbara Street, Pasadena, CA 91101, USA}

\author[0000-0002-7541-9565]{Jonathan Stern}
\affil{Department of Physics and Astronomy and CIERA, Northwestern University, Evanston, IL 60208, USA}

\author{Andy D. Goulding}
\affil{Department of Astrophysical Sciences, 4 Ivy Lane, Princeton University, Princeton, NJ 08544, USA}

\author{Jenny E. Greene}
\affil{Department of Astrophysical Sciences, 4 Ivy Lane, Princeton University, Princeton, NJ 08544, USA}

\author[0000-0001-6100-6869]{Nadia L. Zakamska}
\affil{Department of Physics and Astronomy, Bloomberg Center, Johns Hopkins University, Baltimore MD 21218, USA}

\author[0000-0003-2830-0913]{Rachael M. Alexandroff}
\affil{Dunlap Institute for Astronomy and Astrophysics, University of Toronto, 50 St. George Street, Toronto, Ontario, Canada M5S 3H4}

\author[0000-0001-8813-4182]{Hsiao-Wen Chen}
\affil{Department of Astronomy \& Astrophysics, The University of Chicago, 5640 S Ellis Ave., Chicago, IL 60637, USA}

\correspondingauthor{Sean D. Johnson}
\email{sdj@astro.princeton.edu}

\begin{abstract}
Galactic-scale winds driven by active galactic nuclei (AGN) are often invoked to suppress star formation in galaxy evolution models, but the mechanisms driving these outflows are hotly debated. Two key AGN feedback models are (1) radiation pressure accelerating cool gas and (2) a hot outflowing wind entraining the ISM. Highly ionized emission-line diagnostics represent a powerful means of differentiating these scenarios because of their sensitivity to the expected compression of the ISM clouds by the hot wind. Here, we report the first spatially resolved UV emission spectroscopy of a prototypical (radio-quiet) quasar-driven superwind \SDJ{around the obscured quasar SDSS\,J1356+1026 at $z=0.123$}. We observe ratios of O\,VI/C\,IV, N\,V/C\,IV, and C\,IV/He\,II that are remarkably similar for outflowing gas clouds $\lesssim 100$ pc and $\approx 10$ kpc from the nucleus. Such similarity is expected for clouds with AGN radiation pressure dominated dynamics. Comparing the observed line emission to models of clouds in balance with radiation pressure and/or a hot wind, we rule out the presence of a dynamically important hot wind and constrain the ratio of hot gas pressure to radiation pressure to $P_{\rm hot}/P_{\rm rad} \lesssim 0.25$ both at $\lesssim 100$ pc and $\approx 10$ kpc from the nucleus. Moreover, the predictions of the radiation pressure confined cloud models that best fit observed UV line ratios are consistent with the observed diffuse X-ray spectrum. These results indicate that this AGN superwind is driven by radiation pressure or was driven by a hot wind that has since dissipated despite on-going AGN activity.

\end{abstract}

\section{Introduction} \label{sec:intro}
Modern models of galaxy evolution often invoke powerful feedback from accreting supermassive black holes (SMBH) in galactic nuclei in order to suppress star formation in massive galaxies \citep[for reviews, see][]{Kormendy:2013, Heckman:2014, Somerville:2015}. Direct observations of powerful circum-nuclear active galactic nuclei (AGN) driven winds demonstrate that SMBH feedback is in principle possible, but the physical mechanisms that couple energy and momentum from the nucleus to the interstellar medium (ISM) and surrounding halo gas are fiercely debated \citep[e.g.][]{Morganti:2017, Wylezalek:2018}. Two key models for driving effective, large-scale AGN outflows are (1) direct acceleration of cool gas through radiation pressure \citep[e.g.][]{Murray:2005, Ishibashi:2015, Debuhr:2011, Thompson:2015} and (2) entrainment of the ISM in a hot outflowing wind generated by fast shocks near the nucleus \citep[e.g.][]{Faucher-Giguere:2012, King:2015}. Observations that differentiate between these AGN feedback models are necessary for a more complete understanding of galaxy evolution \citep[e.g.][]{Krumholz:2017}.

Over the last decade, surveys of outflowing gas around luminous (radio-quiet) AGN demonstrate that multi-phase and kinematically disturbed outflows are nearly ubiquitous both near the nucleus \citep[$<1$ kpc  e.g.][]{Feruglio:2010, Zakamska:2014} and on galactic scales of $\approx 10$ kpc \citep[e.g.][]{Greene:2011, Greene:2012, Zubovas:2012, Liu:2013, Liu:2013a, Liu:2014, Harrison:2014, Rupke:2017, Sun:2017, Fischer:2018, Husemann:2019, Jarvis:2019}. However, more in-depth follow-up observations are needed to make definitive statements about the physical conditions in the ongoing outflows, let alone the physical mechanisms that drove or continue to drive them.

Emission line ratios of highly ionized species can diagnose the physical conditions in quasar outflows. These line ratios are sensitive to the ionization level of the H\,II gas, which in turn depends on the dominant pressure source applied to the illuminated surface of the clouds \citep[][]{Stern:2016}. If quasar radiation pressure is the dominant pressure source then the thermal gas pressure at the ionization front roughly equals the incident radiation pressure, implying an ionization parameter of $U{\sim}0.03-0.1$. Moreover, in radiation-pressure-dominated clouds, the H\,II layer has a characteristic density profile and spectral signature in highly-ionized lines \citep[][]{Stern:2014b, Stern:2014a, Baskin:2014a, Baskin:2014b, Bianchi:2019}.  If another pressure mechanism dominates -- such as the hot wind -- then the gas will have a higher pressure/density and thus a lower ionization state \citep[e.g.][]{Dopita:2002, Stern:2016}. In addition, highly ionized lines can differentiate AGN photoionized gas \citep[][]{Groves:2004} from shocks \citep[][]{Allen:2008}. Consequently, spectroscopy of emission lines such as O\,VI, N\,V, and C\,IV in the rest-UV represents a sensitive means of testing AGN feedback mechanisms and constraining the elusive hot wind phase.

\begin{figure*}
\gridline{\fig{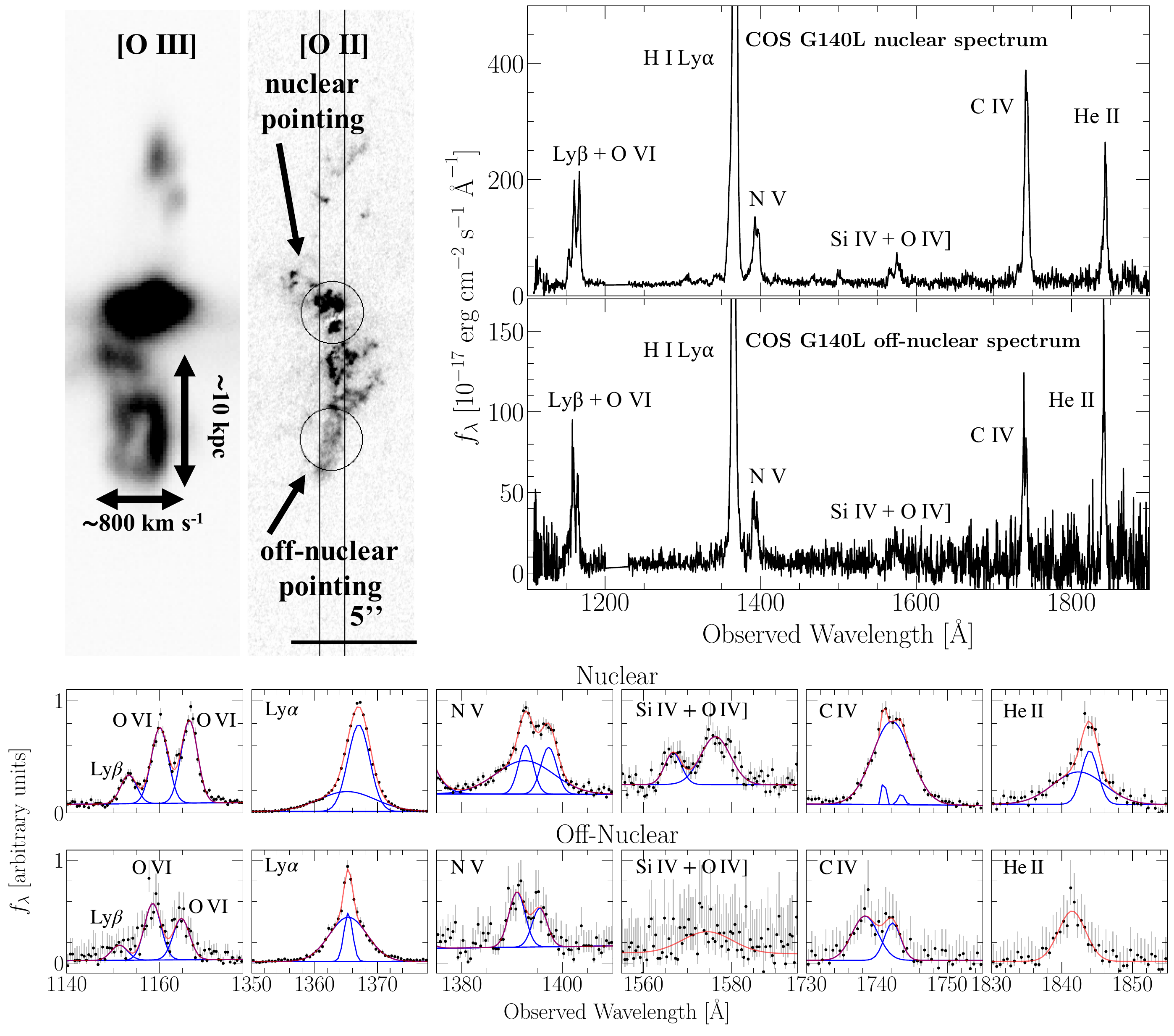}{1.0\textwidth}{}}
\caption{Summary of observations of SDSS\,J1356$+$1026, a prototypical quasar-driven superwind at $z=0.123$. The {\it top left} panel displays a 2D longslit spectrum from Magellan centered on the [O\,III] $\lambda$5007 line showing extended and kinematically disturbed ($\Delta v \approx 800$ \kms) [O\,III] emission $\approx 10$ kpc South of the nucleus. The {\it top middle} panel displays a high resolution {\it HST} image of the [O\,II] $\lambda$3727 line emission formed by subtracting the F438W and F621W bands on the same spatial scale as the 2D spectrum. \SDJ{The emission-line map is dominated by [O\,II] but may also contain non-negligible contribution from scattered light \citep[e.g.][]{Dempsey:2018}.}. The locations of the COS aperture and \SDJ{1'' wide} Magellan longslit are overlaid to scale on the {\it HST} image for both the nuclear and off-nuclear pointings. The {\it top two} panels on the {\it right} display the COS spectra from the nuclear and off-nuclear pointings. The {\it bottom} panels show zoom-in emission-line spectra for the nuclear ({\it top} panels) and the off-nuclear ({\it bottom} panels) COS pointings \SDJ{with emission-lines labelled}. The best-fit emission model is shown as a red line and individual Gaussians are shown in blue.
\label{figure:J1356}}
\end{figure*}

Here, we present the first sensitive and spatially resolved UV spectra of a prototypical quasar-driven superwind at low redshift, SDSS\,J135646.10+102609.0 (SDSS\,J1356+1026). SDSS\,J1356+1026 is a radio-quiet, obscured quasar at redshift $z=0.123$ driving a prototypical superwind on galactic scales \citep[][]{Greene:2012}. The AGN has an estimated bolometric luminosity of $L_{\rm bol} \approx 2 \times 10^{46} \lum$ and a black hole mass of $M_\bullet \sim 10^8\,{\rm M}_\odot$ \citep[][]{Sun:2014}.
The outflow is characterized by kinematically disturbed ionized gas \SDJ{with double-peaked velocity structure} (full width $\gtrsim 800$ \kms\ in projection, inconsistent with bound gravitational motion) observed in [O\,III] emission at $\approx 10$ kpc from the nucleus that can be modelled as outflowing shells \citep[see Figure \ref{figure:J1356};][]{Greene:2012}. Soft X-ray emission that is detected at the location of the extended outflow can be explained either by the presence of a hot wind or by photoionized line emission \citep[][]{Greene:2014, Foord:2020}.

Throughout, we adopt a flat $\Lambda$CDM cosmology with $\Omega_{\rm m}=0.3$, $\Omega_\Lambda = 0.7$, and $H_0 = 70\ {\rm km\,s^{-1}\ Mpc^{-1}}$.

\section{Observations and Data Reduction} \label{sec:obs}

\subsection{HST COS Data Reduction and Measurements}
We obtained sensitive, spatially resolved FUV emission-line spectra of SDSS\,J1356+1026 with the Cosmic Origins Spectrograph \citep[COS;][]{Green:2012} on board the {\it Hubble Space Telescope} ({\it HST}) both with a nuclear pointing \SDJ{on 2019-06-12 and 2019-06-14} (4 orbits; \SDJ{10.39 sec of exposure; OBSIDs: LDHV02010, LDHV01010; PI: Johnson, PID: 15280}) and an off-nuclear pointing \SDJ{on 2018-05-20} (1 orbit; \SDJ{2.16 ksec of exposure; OBSID: LDHV03010}). The off-nuclear pointing is centered at the location of the extended outflow observed in [O\,III], 4.8 arcsec (10.6 kpc) South and 0.6 arcsec (1.6 kpc) West of the nucleus as shown in the middle panel of  Figure~\ref{figure:J1356}. The COS G140L grating spectra cover key high ionization lines such as the O\,VI $\lambda\lambda$1031/1037, N\,V $\lambda\lambda$1238/1242, C\,IV $\lambda\lambda$1548/1550, and Si\,IV $\lambda\lambda1394/1403$ doublets as well as He\,II $\lambda 1640$  and the O\,IV] multiplet at $\lambda1400$ (blended with Si\,IV).

We calibrated the COS spectra using the \calcos{} pipeline version 3.3.5. Because \calcos{} is optimized for point sources, we enlarged the spectral extraction aperture size to $\approx49$ pixels (5.4 arcsec) from the default. Our chosen extraction aperture includes at least 98\% of the total flux under the O VI, N V, and C IV lines, and the line ratios are robust to aperture size changes at the level of $\pm 10\%$. We combined the extracted individual exposures for each pointing into exposure time weighted final spectra after masking bad  pixels. Figure~\ref{figure:J1356} shows the final nuclear and off-nuclear spectra in the top two panels on the right.

The nuclear and off-nuclear UV spectra from COS shown in Figure~\ref{figure:J1356} exhibit emission in H\,I \lyb, O\,VI, H\,I \lya, N\,V, Si\,IV$+$O\,IV], C\,IV, and He\,II $\lambda 1640$. The observed line ratios in the off-nuclear spectrum are strikingly similar to those in the nuclear spectrum. To measure the strengths of the emission lines and quantify this similarity, we fit the spectral region around each line with a linear continuum model and Gaussian emission components as shown in the zoom-in plots in Figure~\ref{figure:J1356}. Some features required multiple Gaussians to achieve a good fit.
Due to low signal-to-noise in the Si\,IV+O\,IV] emission line for the off-nuclear spectrum, we fixed the width of the fit to that from a single Gaussian fit to the same region in the nuclear spectrum. For the same reason, we fixed the width of the Ly$\beta$ off-nuclear Gaussian to be the same as that of the off-nuclear O\,VI lines.
The line measurements are summarized in Table~\ref{tab:lines} after 
Milky Way foreground extinction corrections based on \cite{Schlafly:2011} and \cite{Fitzpatrick:2007}. The line flux uncertainties include systematic errors of 10$-$20\% based on flux measurement variations using different continuum models and non-parametric measurements. The $\approx 1''$ spatial extent of the emitting gas in the dispersion direction is expected to produce spectral resolution of $\approx 600-1000$ \kms\ from 1800-1100 \AA, consistent with the observed line and doublet full-width-at-half-maxima (FWHM).

\subsection{HST WFC3 Imaging}

To assess the morphological structure of the ionized gas around SDSS\,J1356+1026, we constructed an [O\,II] line-map using available wide and medium band {\it HST} images from WFC3+UVIS in the F438W (PI: Comerford, PID: 12754) and F621M (PI: Greene, PID:13944) filters. The F438W filter includes both continuum and line emission, predominantly from the [O\,II] doublet, while the F621M filter is free from strong emission lines. To create the emission line map, we started with the default image reductions from STScI and performed an astrometric alignment and flux scaling prior to image subtraction. To align and scale the images, we identified sixteen serendipitous sources residing in the common field-of-view (FOV). We then estimated the optimal translation, rotation, and flux scaling between the two images by simultaneously minimizing the residuals in subtracted $2''\times2''$ cutouts around the serendipitous sources via simulated annealing. The resulting line map is shown in Figure \ref{figure:J1356}.

\subsection{Chandra X-ray Observations}

There are three archival Chandra X-ray observations of SDSS\,J1356+1026 obtained with the Advanced CCD Imaging Spectrometer (ACIS) in ACIS-S mode taken on dates 2012-03-31 (ObsID:13951; FAINT mode), 2016-03-29 (ObsID:17047; VFAINT mode), and 2016-05-19 (Obsid:18826; VFAINT mode). Because of the differences in the observing setup and the sensitivity of ACIS over time, we processed each observation separately but consistently using the Chandra software packages in CIAO v4.11 with calibration files from CALDB 4.8.3 applied using {\tt chandra\_repro}. After removing streak events, bad pixels, pixel randomization, cosmic rays, and flares, the final Level-2 events files consist of Good Time Intervals of 19.8, 34.8, and 42.9 ks respectively.

For our analysis of the diffuse X-ray spectrum, we extracted photons and response files using the {\tt specextract} package from each events file within a region bounded by the 99\% flux contour from the {\it HST} [O\,II] image. To prevent contamination from the central AGN we masked a central circular region of $2''$ radius  (90 and 95\% encircled energy fraction at $E=4.5$ and $2$ keV respectively). We analyzed the resulting X-ray spectra with XSPEC v12.10 using \cite{Cash:1976} statistics and required a minimum of one photon per energy bin.

\begin{deluxetable*}{c|lcrcr}
\tablenum{1}
\tablecaption{Summary of observed UV emission line properties  \label{tab:lines}}
\tablewidth{0pt}
\tablehead{
\colhead{} & \colhead{Line} & \colhead{Flux [$10^{-15} \flux$]} & \colhead{Counts} & \colhead{Centroid [\AA]} & \colhead{FWHM [\kms]}}
\startdata
\multirow{9}{*}{Nuclear} & \lyb & $2.5^{+0.3}_{-0.3}$ & $ 770$ & $1153$ & $1000$ \\ 
& O VI 1031.92/1037.61 & $15^{+1}_{-1}$ & $4480$ & $1160$ & $2700$ \\ 
& H I \lya & $100^{+10}_{-10}$ & $57281$ & $1367$ & $ 1000$ \\ 
& N V 1238.82/1242.80 & $11^{+1}_{-1}$ & $5484$ & $1392$ & $1900$ \\ 
& Si IV 1393.75/1402.77 + O\,IV] 1400 & $3.2^{+0.3}_{-0.3}$ & $ 1030$ & $1573$ & $ 2900$ \\ 
& C IV 1548.19/1550.77 & $28^{+3}_{-3}$ & $3347$ & $1742$ & $1000$ \\ 
& He II 1640.40 & $ 12.1^{+ 1}_{- 1}$ & $1000$ & $1842$ & $600$ \\ 
\hline
\multirow{9}{*}{Off-Nuclear} & \lyb & $0.8^{+0.2}_{-0.2}$ & $  49$ & $1151$ & $1100$ \\ 
& O VI 1031.92/1037.61 & $ 4.9^{+ 0.6}_{- 0.6}$ & $ 310$ & $1161$ & $2500$ \\ 
& H I \lya & $23^{+ 2}_{- 2}$ & $2777$ & $1365$ & $600$ \\ 
& N V 1238.82/1242.80 & $ 2.3^{+ 0.3}_{- 0.3}$ & $ 247$ & $1391$ & $1600$ \\ 
& Si IV 1393.75/1402.77 + O\,IV] 1400& $ 1.5^{+ 0.4}_{- 0.4}$ & $  76$ & $1574$ & $2500$ \\ 
& C IV 1548.19/1550.77 & $ 5.3^{+ 0.7}_{- 0.7}$ & $  138$ & $1739$ & $ 1300 $ \\ 
& He II 1640.40 & $ 5.7^{+ 0.9}_{- 0.9}$ & $ 101$ & $1841$ & $ 700$ \\ 
\enddata

\end{deluxetable*}

\begin{figure*}
\gridline{\fig{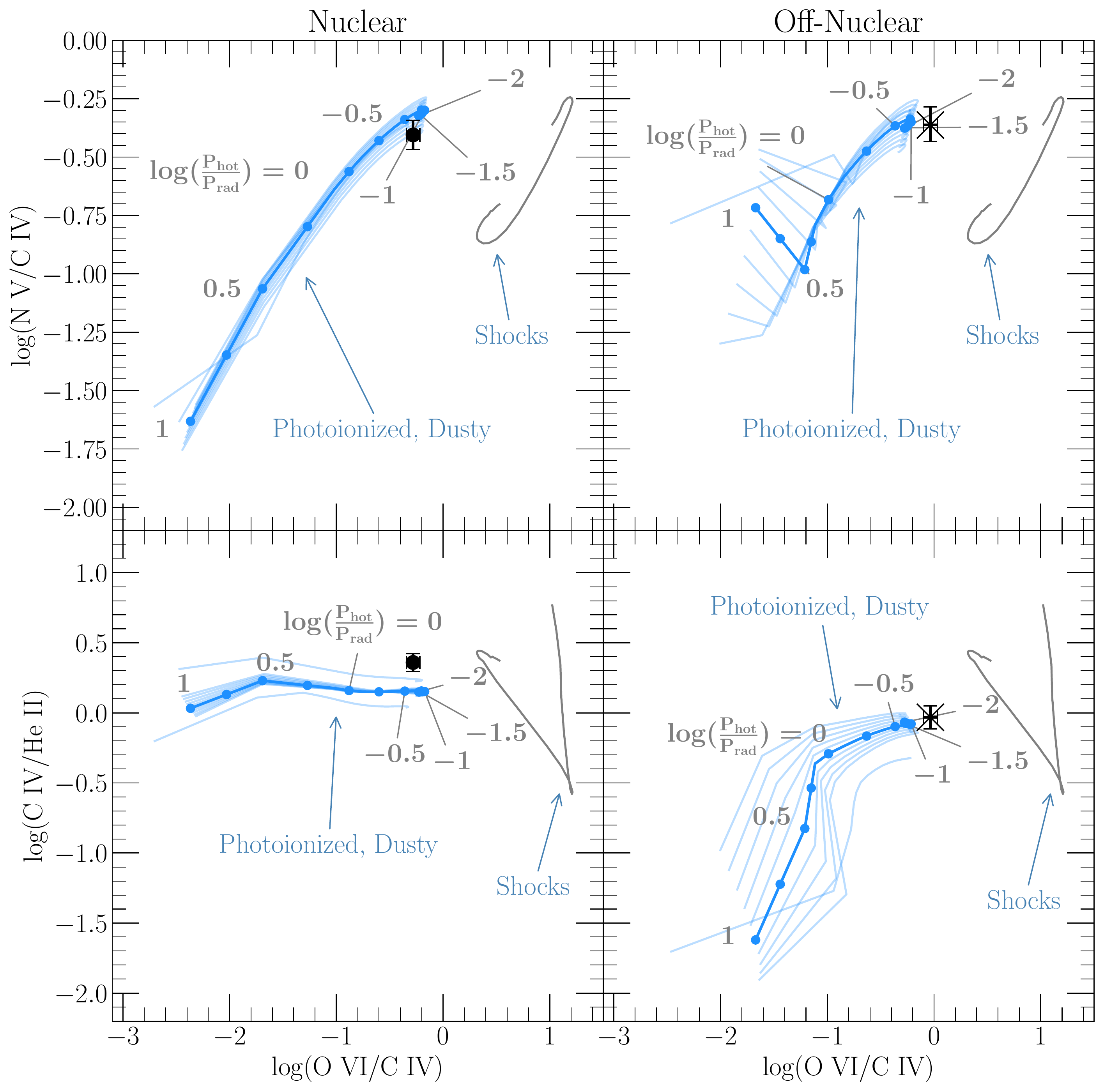}{0.8\textwidth}{}}
\caption{ 
Measured N\,V/C\,IV ({\it top}) and C\,IV/He\,II ({\it bottom}) versus \SDJ{O\,VI/C\,IV} emission line ratios for the nuclear and off-nuclear pointings compared to predictions from shock models and AGN photoionized hydrostatic models. Shock models with shock velocities $>200$ \kms\  \citep[][]{Allen:2008} are shown in grey and are inconsistent with the observed line ratios. Shock models with lower velocities cannot reproduce the optical line ratios observed in \cite{Greene:2012}. Hydrostatic AGN photoionization models (blue) include dust grains. The models span $P_{\rm hot}/P_{\rm rad}=0.01 - 10$ in steps of $0.25$ dex as marked. Different lines mark different assumptions on the metallicity, spectral slope, and distance of the clouds as described in Section~\ref{section:hydro}. The predicted line ratios approach asymptotic values as $P_{\rm hot} / P_{\rm rad} \rightarrow 0$, corresponding to the radiation pressure dominated limit. The observed ratios are most consistent with gas clouds in the radiation pressure dominated limit, ruling out the current presence of a dynamically important hot wind.
\label{fig:ratios}}
\end{figure*}

\section{Discussion} \label{sec:disc}

The observed emission line ratios of N\,V/C\,IV, C\,IV/He\,II, and \SDJ{O\,VI/C\,IV} are shown in Figure~\ref{fig:ratios}, and are strikingly similar for the nuclear and off-nuclear observations, despite the factor of $\gtrsim10^4$ difference in ionizing flux. For example, the nuclear N\,V to C\,IV ratio of $0.40\pm 0.05$ is consistent with the off-nuclear value of $0.43\pm 0.08$. This suggests that the density of the emitting clouds has the same $\propto r^{-2}$ radial dependence as the quasar radiation field, as expected if radiation pressure dominates.

To better understand the physical conditions of the emitting clouds, we compare the observed line ratios in Figure \ref{fig:ratios} to models including radiative shocks \citep[plus photoionized precursor;][]{Allen:2008} and AGN photoionized models calculated with \cloudy{} version 17.01 \citep{Ferland:2017}.
Shocks with velocities greater than $500$ \kms\ can reproduce the observed emission line ratios of [O\,III]/H$\beta \approx 10$ at the nucleus and location of the off-nuclear pointing from \cite{Greene:2012}. However, at these velocities, the shock models significantly over-predict the observed highly ionized line ratios (e.g. \SDJ{O\,VI/C\,IV}) as shown in Figure \ref{fig:ratios}. \SDJ{We therefore conclude that AGN photoionization dominates over shock ionization by a large factor in the observed regions of SDSS\,J1356+1026, though we caution that shocks may still be present in the system \citep[e.g.][]{Zakamska:2014}.}

\subsection{\SDJ{Constant density models}}

The observed O\,VI/C\,IV ratios can be reproduced by \JS{models of AGN photoionized gas clouds with uniform density}, $\approx 1{-}2\times$ solar metallicity, and ionization parameter of $\log U\approx-0.7$  \citep[][]{Groves:2004}, implying $n_{\rm H}\approx 3\times10^4\ {\rm cm}^{-3}$ for $r\lesssim100$ pc and $\approx2\ {\rm cm}^{-3}$ for $r\approx 10$ kpc assuming an AGN bolometric luminosity $L_{\rm bol} = 2 \times 10^{46} \lum$ \citep[][]{Sun:2014} and nominal distances of 100 pc and 10 kpc for the nuclear and off-nuclear pointings. Such AGN photoionized gas clouds exhibit an equilibrium temperature of $\approx 10^{4}$ K implying gas pressures of $P_{\rm gas} \approx 10^{8}$ and $\approx 5\times10^{4}$ K cm$^{-3}$ for the nuclear and off-nuclear pointings respectively, $\gtrsim7\times$ less than the \JS{pressure in the incident radiation, $P_{\rm rad} \equiv L/(4\pi r^2 c) \approx 4\times10^9$ and $4\times10^5$ $\rm K\,cm^{-3}$ respectively}. Consequently, \JS{neglecting the effect of radiation pressure on the structure of the ionized gas is not justified (e.g.\ \cite{Dopita:2002})}.

\begin{figure*}
\gridline{\fig{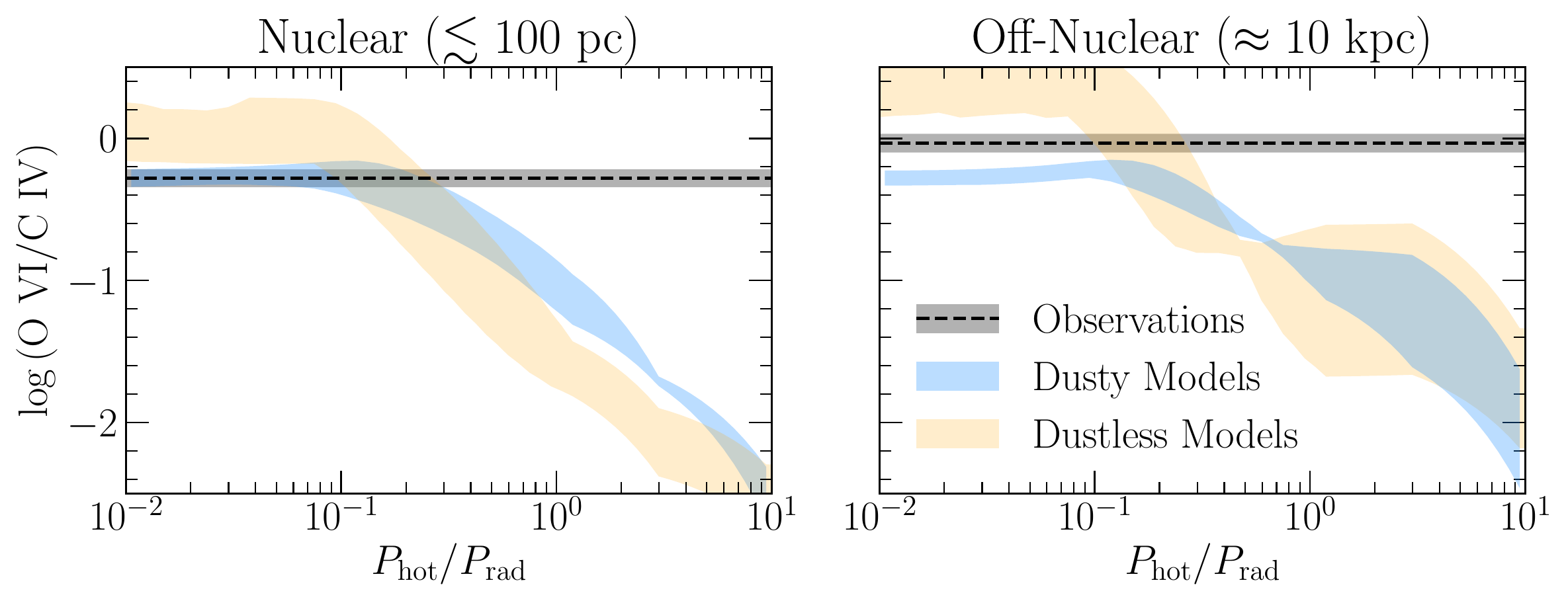}{1\textwidth}{}}
\caption{Predicted O\,VI/C\,IV emission line ratios as a function $P_{\rm hot}/P_{\rm rad}$ for the dusty (dustless) hydrostatic AGN photoionized models in blue (orange) with thickness denoting the predicted range including model uncertainty discussed in the text. The observed line ratios are shown as black dashed lines with grey regions denoting uncertainties. The observed line ratios for both pointings require $P_{\rm hot}/P_{\rm rad} \lesssim 0.25$, ruling out the presence of a dynamically important hot wind.
\label{fig:money}}
\end{figure*}
\subsection{Hydrostatic models including AGN radiation pressure}\label{section:hydro}

\JS{To account for the effects of radiation pressure, we employ hydrostatic models of ionized clouds, in which the pressure at the illuminated surface is set by the thermal pressure of the ambient hot gas, $P_{\rm hot}$, while the momentum transferred to the gas via the absorption of radiation is balanced by a thermal pressure gradient within the ionized cloud. Such models are sometimes referred to as ``constant total pressure'' models, since the sum of the thermal gas pressure and the remaining pressure in the absorbed radiation is constant throughout the slab.\footnote{These models are calculated in \cloudy{} using the ``constant pressure'' option, though they should not be confused with constant gas pressure models in which $P_{\rm gas}$ is held constant.} Confinement on the shielded side is assumed to be provided by the neutral/molecular gas beyond the ionization front. As discussed in  \cite{Dopita:2002} and \cite{Stern:2016}, the structure of the ionized cloud depends qualitatively on whether hot gas or radiation is the dominant pressure source. If $P_{\rm hot}\gg P_{\rm rad}$ then radiation pressure is negligible and the gas pressure is roughly constant throughout the cloud with $P_{\rm gas}\approx P_{\rm hot}$, resulting in a roughly uniform density ionized layer. In contrast if $P_{\rm rad} \gg P_{\rm hot}$ then the gas pressure increases significantly with depth into the cloud, from $P_{\rm gas}=P_{\rm hot}$ at the illuminated surface to $P_{\rm gas} \approx P_{\rm rad}$ near the ionization front (see fig.~1 in \cite{Stern:2016}). In this case the cloud is \textit{Radiation Pressure Confined (RPC)}, and density and hence ionization parameter are a function of depth so that highly ionized lines arise primarily from outer layers while lower ionization lines arise from deeper layers closer to the ionization front. In RPC clouds, the predicted line ratios are independent of $P_{\rm hot}$ and exhibit unique spectral signatures. Consequently, highly ionized FUV emission line observations of AGN outflows can serve as effective barometers that enable inferences into whether radiation pressure or a hot wind determine the dynamics of the cool-warm component of AGN driven outflows.}

\label{sec:xray}
\begin{figure*}
\includegraphics[width=\textwidth]{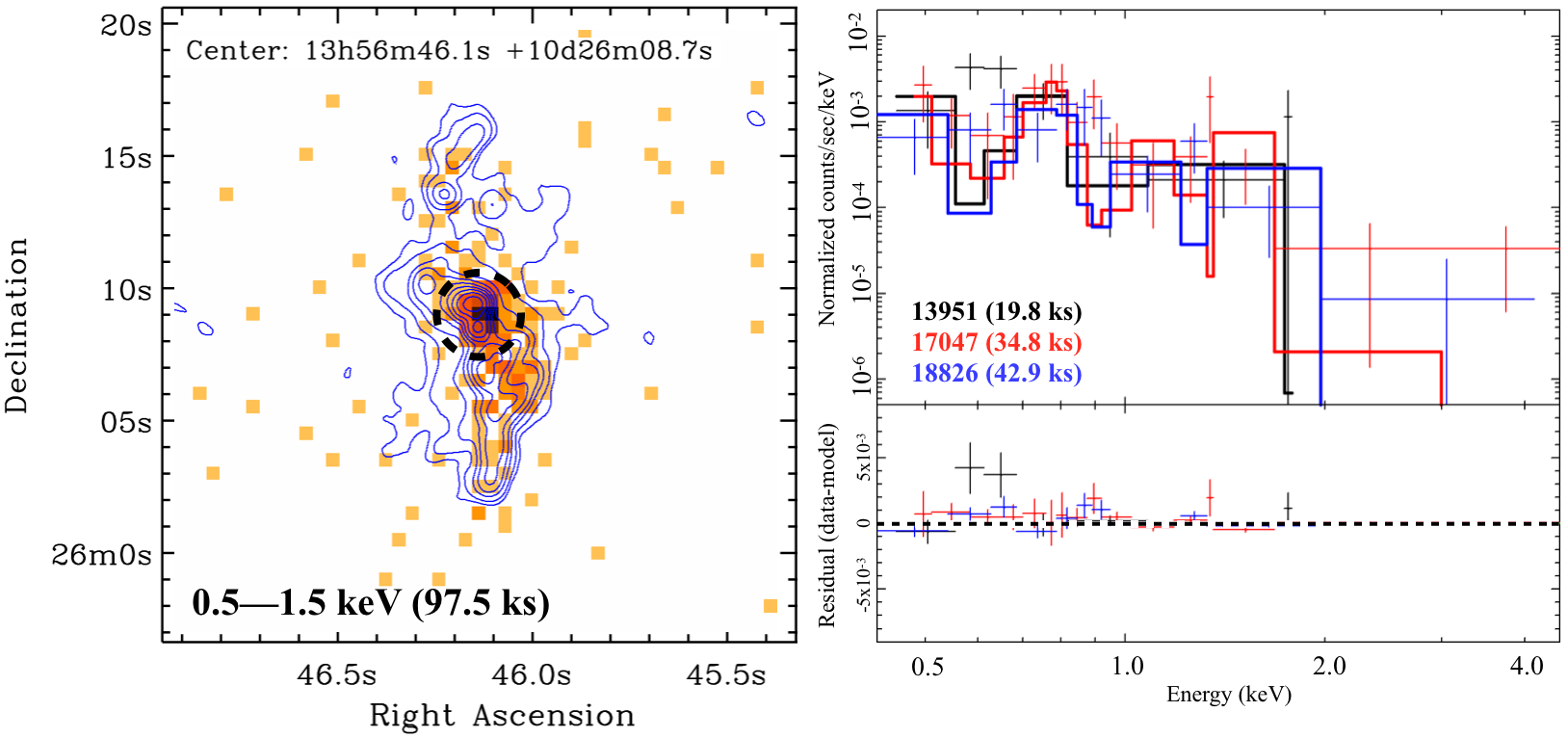}
\caption{ {\it Left:} Soft band (0.5--1.5 keV) Chandra X-ray image produced from all three Chandra programs. Contours of emission line regions from HST imaging are overlaid in blue, the outermost contour represents 99\% of the total line flux, and is used to extract X-ray photons for our spectral analysis shown on the {\it right}. The circle marks the region masked in the spectral extraction to prevent contamination from the nucleus. {\it Right:} X-ray spectra in counts/sec/keV of the diffuse X-ray component extracted from the line emitting region. X-ray data from all three Chandra ObsIDs are shown simultaneously in different colors with a single {\sc cloudy} photoionization model (solid line). Residuals between the X-ray data and the models are shown on the bottom.
\label{fig:xray}}
\end{figure*}

\JS{
We used \cloudy{} to calculate the structure of hydrostatic H\,II regions photoionized by AGN over a wide range of relative pressures ($0.01 < P_{\rm hot}/P_{\rm rad}< 10$). Other model parameters include the dust content of the gas, gas metallicity, the ionizing spectral slope $\alpha$ ($L_\nu\propto \nu^{-\alpha}$) between 1 Rydberg and 2 keV, and the distance to the unresolved nucleus pointing. For the purposes of this letter, these are nuisance parameters. We ran a grid of models with $\alpha=1.6 - 2$, metallicity in the range $1-2\,{\rm Z}_{\odot}$, and Milky Way ISM abundances and dust content/depletion  \citep[][]{Draine:2011}. Because dust may be destroyed in the AGN outflow, we also consider dust-free models which are discussed in the Appendix. Our conclusions are robust to a wide range in choice of these parameters. We assume an AGN
bolometric luminosity of $L_{\rm bol} = 2 \times 10^{46}$ erg s$^{-1}$, and distances of $10-100$ pc for the nuclear pointing and a distance of $10$ kpc for the off-nuclear pointing. The predicted line ratios from the models are shown in Figure \ref{fig:ratios}. Dots connected by thick lines denote the model predictions for different $P_{\rm hot}/P_{\rm rad}$, for an assumed $\alpha=1.6$, $1.5\, {\rm Z}_\odot$, and a distance of 100 pc (10 kpc) for the nuclear (off-nuclear) pointing. Thin lines denote predicted line ratios for other choices of these three parameters. \SDJ{Because the cloud models are in the optically thick limit the size and total columns of the emitting regions are not not free parameters. Nevertheless, we ensured that the emitting cloud sizes are smaller than the COS aperture and the corresponding total hydrogen columns range from $N({\rm H})=10^{20}-10^{21.7}$ cm$^{-2}$.}

As $P_{\rm hot}/P_{\rm rad} \rightarrow 0$, the predicted line ratios approach asymptotic values, as expected in the radiation pressure dominated limit in which the line ratios are independent of $P_{\rm hot}$ \citep{Dopita:2002, Stern:2014a}. In the dusty models shown in Figure \ref{fig:ratios}, the uncertainty in the predicted line ratios given our assumed range of metallicity, spectral slope and cloud distance is merely $\lesssim 0.1$ dex. All observed line ratios are within $\approx0.1$ dex of the predictions for radiation pressure dominated clouds for both the nuclear pointing and the off-nuclear pointing. The hot gas pressure dominated models with $P_{\rm hot}>P_{\rm rad}$ under-predict O\,VI/C\,IV by an order-of-magnitude and underpredict the observed N\,V/C\,IV by a factor of $\approx3$. The observed line ratios thus strongly disfavor the hot gas pressure dominated models. 
}

\subsection{Any hot wind is currently dynamically unimportant}
The observed N\,V/C\,IV, C\,IV/He\,II, and \SDJ{O\,VI/C\,IV} line ratios for both the nuclear and off-nuclear pointings are 
consistent with the hydrostatic model predictions in the radiation pressure dominated regime ($P_{\rm hot} < P_{\rm rad}$) as shown in Figure \ref{fig:ratios}. To quantify the limit on the presence of a hot wind component from the UV spectra, Figure \ref{fig:money} displays the observed nuclear and off-nuclear O\,VI/C\,IV ratios compared to hydrostatic photoionization model predictions as a function of $P_{\rm hot}/P_{\rm rad}$. \JS{The thickness of the colored lines denotes the uncertainty in the prediction due to the uncertainty in the nuisance parameters mentioned in the previous section}. \SDJ{The observed line ratios for the nucleus fall within the uncertainty range for the dusty model while the off-nuclear ratios fall between the dusty and dust-free models\footnote{\SDJ{This suggests intermediate dust content which we will explore in future work.}}.} \SDJ{In both cases, the observed line ratios require $P_{\rm hot} \lesssim 0.25 P_{\rm rad}$.}
We therefore conclude that the outflowing, UV emitting clouds on narrow-line region scales of $\lesssim 100$ pc and on galactic scales of $\approx 10$ kpc  are not currently entrained in a dynamically important hot wind. Using the estimated radiation pressure at the fiducial distances, we place limits on the current pressure from any hot wind of $P_{\rm hot} < 10^9$ and $10^5$ $\rm K\,cm^{-3}$ for the nuclear and off-nuclear pointings, respectively.

\subsection{The extended X-ray emission is consistent with photoionized line emission}

{\it Chandra} observations of SDSS\,J1356+1026 show that the extended outflow of SDSS\,J1356+1026 emits soft X-rays which can be explained by shocks induced by a hot wind \cite[e.g.][]{Choi:2014, Nims:2015} or by AGN photoionized line emission \citep[e.g.][]{Sambruna:2001}. The \SDJ{diffuse, extended} X-ray emission from SDSS\,J1356+1026 \SDJ{is coincident with} [O\,II] emission (see  Figure~\ref{fig:xray}), consistent with either scenario. \AG{This diffuse component is characterized by low-energy X-ray emission with $E\lesssim 2$~keV. At harder energies, there are only 7 X-ray photons with $E \sim 2$--7~keV within the [O\,II] bounded region, fully consistent with the 6.5$\pm$0.9 counts expected from the X-ray background in that area in the three combined {\it Chandra} exposures.}

\AG{To test the two scenarios for the origins of the diffuse X-ray emission, we constructed a \cloudy{} model of diffuse X-ray emission produced by AGN photoionized gas assuming the X-ray emitting layer is dustless since grains will be destroyed by sputtering in X-ray emitting layers, even if the lower-ionization layers which produce the UV emission are dusty \citep{Stern:2014a}. This \cloudy{} model is consistent with the observed X-ray spectra with a Cash statistic of $109.4$ for $85$ degrees of freedom (Right panel in Fig.~\ref{fig:xray}) with few significant residuals (lower panel). The diffuse X-ray emission may also be a consequence of shock-heating in a thermally hot plasma. Hence, we also fit the X-ray spectra with an {\sc apec} model in XSPEC and find that it is equally consistent with a low best-fit metallicity of $<2$\% solar, temperature of $T_{\rm X} \sim 0.3$~keV plasma, and no internal absorption. However, given the consistency between the observed soft X-ray spectrum and the model expectations from the radiation pressure dominated cloud emission observed in the UV, we suggest that the diffuse X-ray emission can be fully explained by photoionized line emission. Conclusively differentiating between the two X-ray scenarios will require future X-ray observatories such as Lynx \citep[][]{Gaskin:2018}.}

\section{Summary and conclusions}
\label{sec:concl}
To gain insights into the physical drivers of AGN feedback on galactic scales, we performed spatially resolved UV emission spectroscopy of a prototypical quasar-driven superwind at low-$z$. Despite the large expected difference in ionizing flux, the observed highly ionized UV line ratios on $\approx 10$ kpc scales are similar to those seen near the nucleus ($\lesssim 100$ pc). This similarity is expected if radiation pressure dominates at the illuminated surface of the line-emitting clouds. Indeed, models of clouds confined by radiation pressure from the AGN self-consistently reproduce the observed UV line ratios as well as the spectral and morphological properties of observed diffuse X-ray emission. 

\SDJ{Based on the observed highly ionized UV emission ratios, we rule out the presence of a dynamically important hot wind phase and place an upper limit on the pressure a hot wind may impart to the UV emitting clouds at $\lesssim10$ kpc. This upper limit is an order-of-magnitude lower than recent estimates based on tentative detections ($3-4\sigma$) of the Sunyaev-Zel'dovich effect around quasars at $z\sim 2-3$ \citep[][]{Hall:2019, Lacy:2019}. This tension can be reconciled if AGN feedback varies significantly from object-to-object (e.g. due to luminosity or redshift), if the hot wind is no longer co-spatial with the UV emitting clouds, or if the hot wind has expanded adiabatically and is no longer a dominant pressure source at this stage in the evolution of SDSS\,J1356+1026. \SDJ{While a hot wind component of the outflow may therefore still exist, the observed highly ionized emission-line ratios indicate that the combined gas pressure and ram pressure from any hot gas are subdominant to the radiation pressure and {\it hence do not confine or provide on-going acceleration to the outflowing, UV emitting clouds}}. The observed AGN outflow is therefore most likely the result of radiation pressure or a hot wind that accelerated the gas at earlier epochs and has since vented or cooled despite on-going AGN activity, placing novel and stringent constraints on AGN feedback models.}

\acknowledgments
We are grateful to T. Heckman and E. Schneider for fruitful discussions that influenced this project. We thank the anonymous referee for a constructive and thorough review which strengthened the paper. S. Johnson acknowledges support from NASA through Hubble Fellowship grant (HST-HF2-51375.001-A) awarded by the Space Telescope Science Institute, which is operated by the Association of Universities for Research in Astronomy, Inc., for NASA, under contract NAS5-26555. Based on observations made with the NASA/ESA Hubble Space Telescope through programs GO-12754, GO-13944, and GO-15280 and retrieved from the HST data archive at the Space Telescope Science Institute. Support for this work was provided by NASA through the grant associated with HST-GO-15280. NLZ is supported in part by HST-AR-14592. This paper includes data gathered with the 6.5 meter Magellan Telescopes located at Las Campanas Observatory, Chile. The scientific results reported in this article are based in part on observations made by the Chandra X-ray Observatory.

\appendix
\section{Dustless Models}
\begin{figure*}
\gridline{\fig{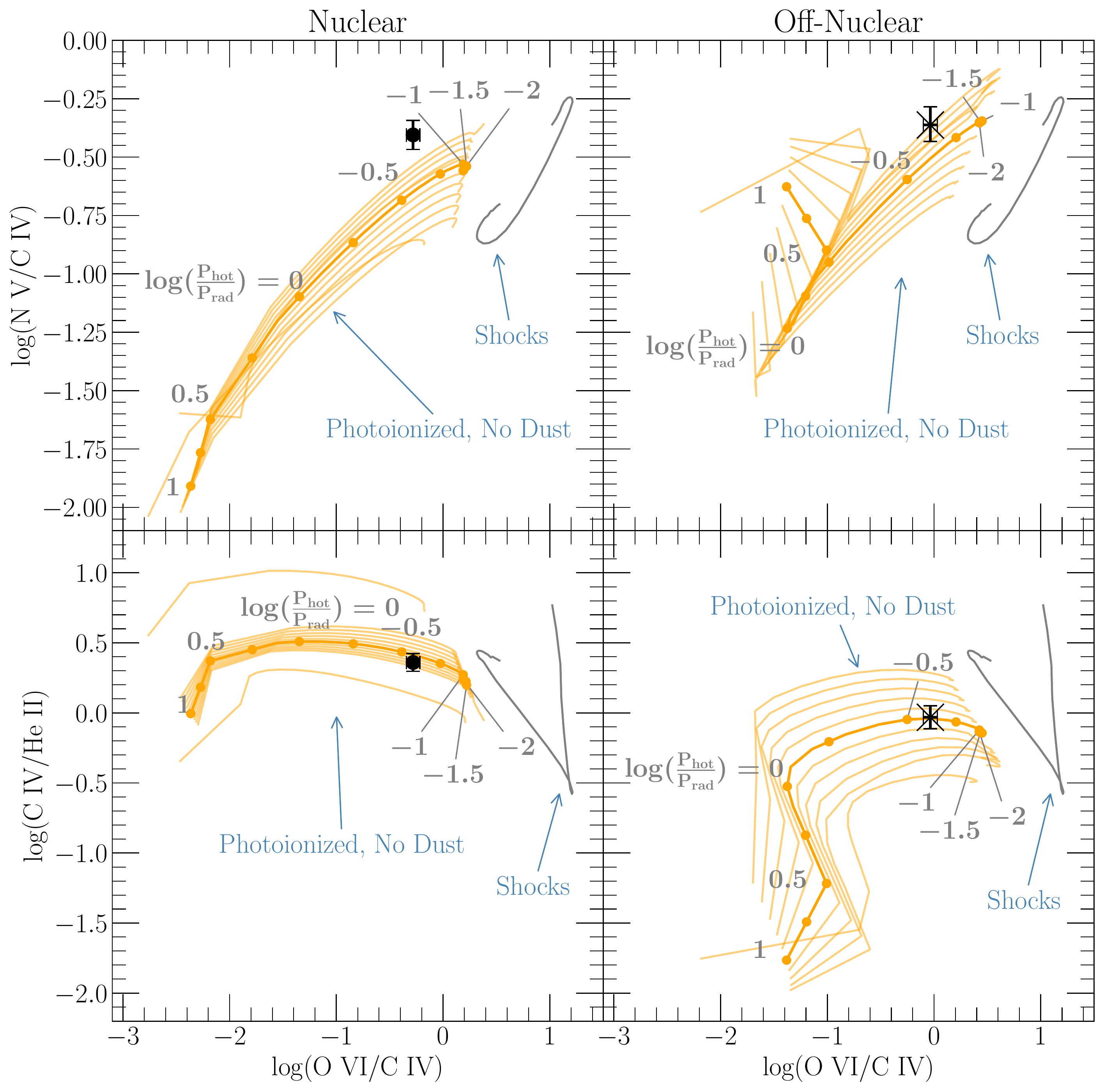}{0.8\textwidth}{}}
\caption{ 
\SDJ{Measured N\,V/C\,IV ({\it top}) and C\,IV/He\,II ({\it bottom}) versus \SDJ{O\,VI/C\,IV} emission line ratios for the nuclear and off-nuclear pointings compared to predictions from shock models and AGN photoionized hydrostatic models, in the same format as Figure~\ref{fig:ratios}. Hydrostatic AGN photoionization models (orange) are dustless. Like the dusty models, the dustless models are most consistent with the observed ratios in the radiation pressure dominated limit, ruling out the current presence of a dynamically important hot wind.}
\label{fig:ratios_nodust}}
\end{figure*}

\SDJ{Dust content of clouds in RPC can change emission-line ratios not only through extinction and gas-phase depletion, but also by altering the thermodynamic properties of the clouds because the dust absorbs radiation pressure. To ensure that our conclusions are robust to dust content of the clouds we ran model grids as described in Section \ref{section:hydro} but with no dust and solar relative abundances \citep[][]{Asplund:2009}. The resulting line ratio predictions are compared to the observed ones in in Figure~\ref{fig:ratios_nodust}. Like with the dusty models, the dustless models are most consistent with the observed ratios in the radiation pressure dominated limit, ruling out the current presence of a dynamically important hot wind.}

 \restartappendixnumbering

\bibliography{main}{}
\bibliographystyle{aasjournal}

\end{document}